\documentclass[apj]{emulateapj}
\usepackage{amsmath}                
\usepackage{amsfonts}               
\usepackage{amssymb}                
\usepackage{epsfig}                 
\usepackage{color}
\usepackage{pifont}
\usepackage[para,online,flushleft]{threeparttable}
\newcommand{\kms}{{~\rm km\; s^{-1}}}

\newcommand{\msun}{~\rm M_{\odot}}
\newcommand{\cm}{{~\rm cm}}
\newcommand{\km}{{~\rm km}}
\newcommand{\s}{{~\rm s}}

\newcommand{\K}{{~\rm K}}
\newcommand{\erg}{{~\rm erg}}

\newcommand{\days}{{~\rm days}}

\newcommand{\revision}[1]{{#1}}
\newcommand{\greentext}[1]{{\textcolor[rgb]{0.00,0.59,0.00}{#1}}}
\newcommand{\redtext}[1]{{\textcolor[rgb]{0.98,0.00,0.00}{#1}}}
\newcommand{\cmark}{\ding{51}}
\newcommand{\xmark}{\ding{55}}

\begin{document}

\title{Explaining the early excess emission of the Type Ia supernova 2018oh by the interaction of the ejecta with disk-originated matter (DOM)}

\author{Naveh Levanon\altaffilmark{1}, Noam Soker\altaffilmark{1,2}}

\altaffiltext{1}{Department of Physics, Technion -- Israel Institute of Technology, Haifa
32000, Israel; nlevanon@campus.technion.ac.il; soker@physics.technion.ac.il}
\altaffiltext{2}{Guangdong Technion Israel Institute of Technology, Shantou 515069, Guangdong Province, China}

\begin{abstract}
We explain the early excess emission of the Type Ia supernova 2018oh by an interaction of the supernova ejecta with disk-originated matter (DOM).
Such DOM can form in the merger process of two white dwarfs (WDs) in the double degenerate scenario of Type Ia supernovae (SNe~Ia).
We find that an ejecta-DOM interaction can fit the early light curve of SN~2018oh better than an ejecta-companion interaction in the single degenerate scenario.
By composing the DOM from two components that were ejected in the merger process with two different velocities, we show that the ejecta-DOM interaction can account for the linear rise in the light curve, while the ejecta-companion interaction predicts too steep a rise.
In addition, the ejecta-DOM interaction does not predict the presence of hydrogen and helium lines in nebular spectra, and hence does not suffer from this major drawback of the ejecta-companion model.
We consider the ejecta-DOM interaction to be the most likely explanation for the early excess emission in SN~2018oh.
By that we show that the double degenerate scenario can account for early excess emission in SNe~Ia.
\end{abstract}

\section{Introduction}
\label{sec:intro}

Five distinguished general binary scenarios claim to bring a white dwarf (WD) or two WDs to explode as a Type Ia supernova (SN~Ia).
We list them by alphabetical order as follows.
The core degenerate (CD) scenario, the double degenerate (DD) scenario, the double-detonation (DDet) scenario, the single degenerate (SD) scenario, and the WD-WD collision (WWC) scenario
(for recent reviews on these five scenarios that include many references to earlier papers and reviews see \citealt{LivioMazzali2018, Soker2018Rev, Wang2018, RuizLapuente2019}).

There are several key observations that constrain one or more scenarios, and there is no scenario that is free of drawbacks.
In some cases observational properties that at first sight seem to belong to one scenario might turn out to be part of another scenario.
Such is the blue excess light in the first days of the explosion that is the subject of this study.
Some studies claimed that it points to the SD scenario, but we show in this letter that the DD scenario can also account for this excess emission.

Several SNe~Ia show early ($\la 5 \days$) excess emission in their light curve.
Notable examples are normal SNe~Ia SN~2012cg \citep{Marioetal2016}, SN~2017cbv \citep{Hosseinzadehetal2017} and SN~2018oh \citep{Shappeeetal2018, Dimitriadisetal2019a}, and the peculiar events iPTF14atg \citep{Caoetal2015} \revision{and MUSSES1604D \citep{Jiang2017}.
A comprehensive list of SNe~Ia with early excess emission appears in \citet{Jiang2018}.}

Many papers prefer the collision of the SN ejecta with a companion in the frame of the SD scenario as the explanation for this excess emission.
In this model the SN ejecta hits a non-degenerate companion and passes through a strong shock wave that heats up the gas.
This post-shock hot gas emits excess UV and blue radiation relative to that of a SN~Ia without an ejecta-companion collision.

There are several alternative explanations for early excess emission.
One alternative is the presence of radioactive nickel in the outskirts of the ejecta.
The nickel heats the ejecta's outer layers leading to excess emission from these layers.
\revision{The presence of heavy elements in the outer layers also reddens the early color curve \citep{Maeda2018}.}
Another alternative is the collision of the ejecta with close circumstellar matter (CSM; \citealt{PiroMorozova2016}).
In particular we have developed an explanation that is based on the collision of the ejecta with disk-originated matter (DOM; \citealt{Levanonetal2015, LevanonSoker2017}).

The ejecta-DOM interaction takes place in the DD scenario.
The more massive WD tidally destroys its companion WD to form an accretion disk.
The accretion disk might blow a bipolar wind (jets) that forms the close CSM that we term DOM.
If explosion occurs shortly, within hours after merger, then the ejecta collides with the DOM to give early excess light.
In an earlier paper we showed that the light curve can be very similar to that expected in the ejecta-companion interaction in the SD scenario, but that we expect neither helium nor hydrogen lines in the spectra.

In the present study we examine the early excess emission of SN~2018oh (ASASSN-18bt).
SN~2018oh is a normal SN~Ia beside its early excess emission.
\cite{Lietal2018} found that the spectral evolution of SN~2018oh is similar to that of a normal SNe~Ia, but that the prominent and persistent carbon absorption features indicate that considerable amount of unburned carbon exists in the ejecta of SN~2018oh.
We will raise the possibility that some of this carbon resides in the DOM.

\cite{Shappeeetal2018} reported and analyzed the early excess emission of SN~2018oh.
They found that the interaction of the ejecta with a non-degenerate companion leads to an abrupt rise and hence cannot adequately explain the initial, slower linear phase.
Their preferred explanation is the presence of $^{56}$Ni in the outskirts of the ejecta, although existing models need tuning.
\revision{\cite{Dimitriadisetal2019a} studied the presence of $0.03\msun$ of $^{56}$Ni on the surface of the ejecta for SN~2018oh, but argue that although this model can explain the early light curve, the expected colors are redder than the observed color curve.
They therefore favored an ejecta-companion collision model.}
However a problem with this collision model is that neither helium nor hydrogen are observed in the nebular phase of SN~2018oh \citep{Tuckeretal2018, Dimitriadisetal2019b}.
\cite{Tuckeretal2018} claim that their results rule out a non-degenerate companion as the explanation for the early excess emission in SN~2018oh.

\cite{Dimitriadisetal2019b} argued that none of the processes they examined can account for the blue excess emission in SN~2018oh.
However, they did not examine the ejecta-DOM collision process.
In this letter we examine the ejecta-DOM interaction as an explanation to the early excess emission of SN~2018oh.
In section \ref{sec:method} we outline the model given in detail in \citet{LevanonSoker2017}.
In section \ref{sec:results} we show our results from fitting DOM models to SN~2018oh, compare to the ejecta-companion collision model and discuss their merits and drawbacks.
We present our conclusions in section \ref{sec:summary}.

\section{Method}
\label{sec:method}

\begin{figure}[ht]
\includegraphics[trim={5cm 10cm 0cm 5cm},clip,scale=0.63]{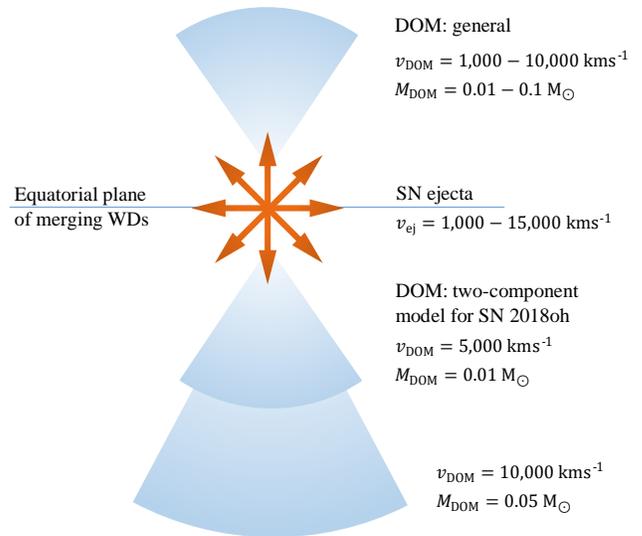}
\caption{\revision{A schematic drawing of the flow structure just after explosion.
The blue regions represent the DOM which has both a partial axial symmetry and a mirror symmetry.
Here the two hemispheres present two cases.
The upper half presents a general DOM structure and the lower half presents the two-component toy model described in section \ref{sec:results}.}}
\label{fig:schematic}
\end{figure}

The full details of the ejecta-DOM collision model are in our previous paper \citep{LevanonSoker2017}.
We recount the main steps and assumptions here.
We assume that prior to the explosion in the DD scenario an accretion disk forms around the more massive WD.
The accretion disk blows a bipolar wind at a velocity of $v_{\rm DOM} \approx 5000 \km \s^{-1}$.
We term the matter expelled from the disk this way DOM.
The DOM has a mass of $M_{\rm DOM} \approx 0.01-0.1 \msun$ spreading in the polar directions over a fraction $f_{\rm DOM} \approx 0.1$ of the full sphere.
\revision{In Fig. \ref{fig:schematic} we present the schematic flow structure.
The upper half of the equatorial plane of the two merging WDs presents the general DOM structure.}

The primary WD explodes at a time $\Delta t_{\rm exp} \approx 10^4 \s$ after disk formation.
The explosion ejecta hits the DOM and shocks it similarly to an ejecta-companion collision \citep{Kasen2010}.
We use an ejecta mass of $M_{\rm ej} = 1.4\msun$ and kinetic energy of $E = 10^{51} \erg$.
We assume the explosion ejecta has an exponential density profile \citep{DwarkadasChevalier1998},
which is maintained after the collision but is compressed by a factor $f_{\rm comp}=1.5$.
The shocked material's initial pressure is ram pressure and it decreases adiabatically with time.
The shocked material has a temperature of $T > 5 \times 10^5 \K$ so that radiation pressure dominates.
We compute the luminosity under the diffusion approximation \citep{Chevalier1992}.
After the photosphere recedes below the strongly-shocked material the luminosity decreases faster as the underlying ejecta layers are only weakly shocked or not at all.
We do not compute the luminosity for this stage.

To fit the luminosity to observations we compute the photospheric radius and effective temperature, assume black-body radiation and filter the flux using the K2 band-pass.
We fit the modelled K2 flux to both the total observed flux and the residual flux after subtracting a power-law model $L \propto t^2$ for the rising light curve.
Since in principle the DOM can contain multiple disconnected parts we also try to fit a two-component DOM model, with material blown at different velocities $v_{\rm DOM}$.

We do not know the exact explosion time of SN~2018oh.
The explosion time we use for the ejecta collision models is the time of first K2 detection, $t_{\rm FD}$, at ${\rm MJD}=8145.1$.
While this cannot be strictly correct, we find the explosion time cannot be much earlier than the first detection time if we assume an interaction model.
\citet{Shappeeetal2018}, \cite{Lietal2018} and \cite{Dimitriadisetal2019a} separately compute different first light times from fitting models of the form $L \propto \left( t - t_0 \right)^\alpha$ to the light curve.
\revision{These first light times are $t_{\rm exp, fit} = t_{\rm FD} - ( 0.19 - 0.45)$~days, i.e., $0.19 - 0.45$~days before first K2 detection.}
Using these times with an ejecta-DOM or an ejecta-companion collision gives a light curve rising too soon by almost as much as $t_{\rm exp, fit}$, as the collision takes place within just a few hours after explosion.
If the source of excess light is an ejecta collision then the extrapolation of first light time using a power-law model is incorrect and explosion must occur just before the first K2 detection for SN~2018oh.
Conversely, if one assumes the expanding fireball model is accurate from the explosion time onwards, then the delay until excess light rules out an interaction model for this event.

\section{Results}
\label{sec:results}

\begin{figure}[ht]
\includegraphics[scale=0.45]{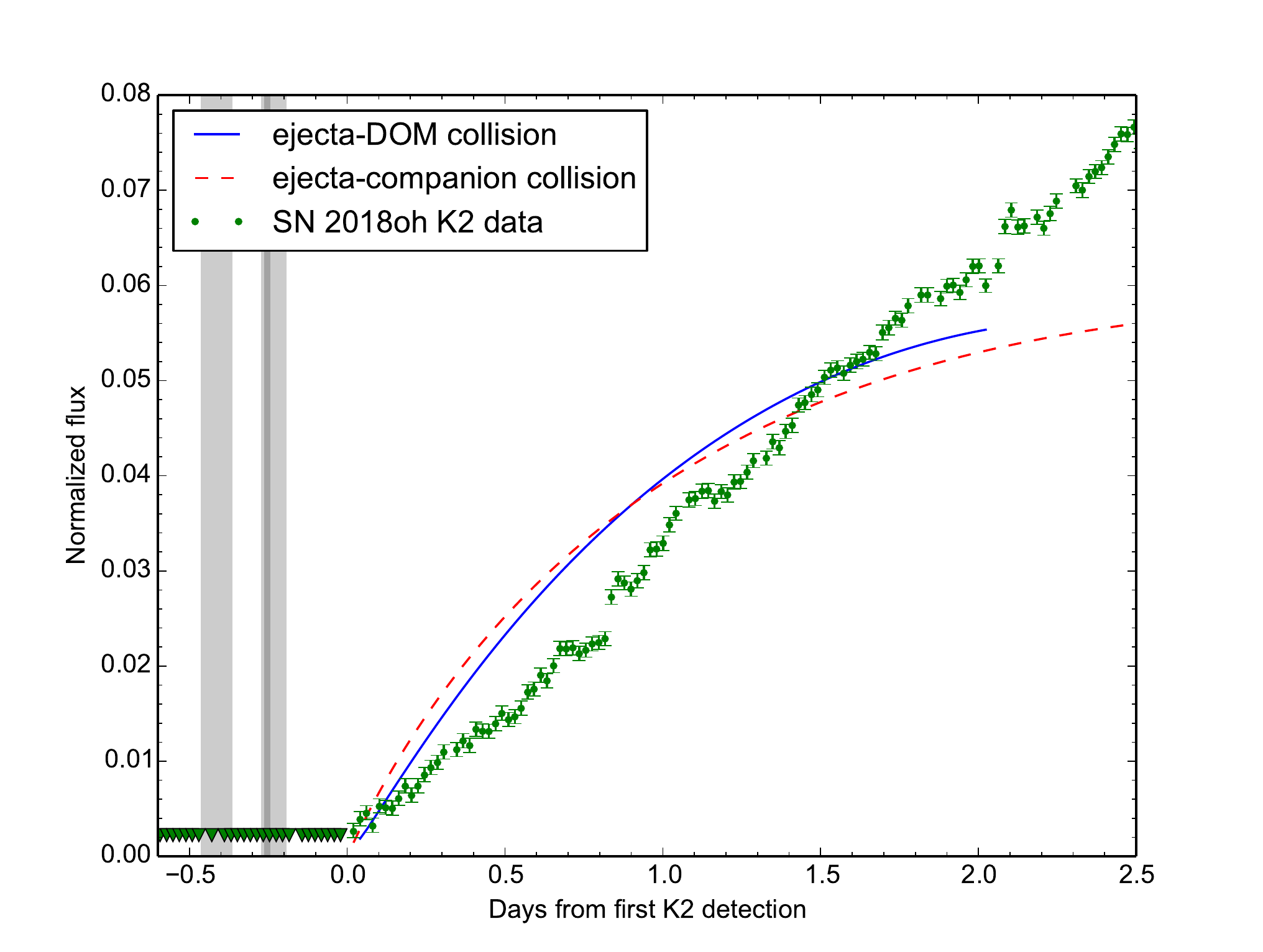}
\caption{Ejecta interaction models for SN~2018oh.
We show the model light curves for an ejecta-DOM collision in solid blue and an ejecta-companion collision in dashed red.
\revision{The models include shock cooling luminosity only and not nickel heating luminosity from the SN itself as the latter is not seen in the first two days.}
We plot the K2 observations of SN~2018oh from \citet{Shappeeetal2018} in green.
Here the DOM includes a single component.
Shaded regions correspond to the first light times calculated by \citet{Dimitriadisetal2019a}, \citet{Lietal2018} and \citet{Shappeeetal2018} from left to right, respectively.}
\label{fig:dom-vs-sd-flux}
\end{figure}

Fig. \ref{fig:dom-vs-sd-flux} shows the ejecta-DOM and ejecta-companion collision models fit to the first days of SN~2018oh.
The shaded regions show the first light times discussed above.
In this figure we present a one-component DOM model.
The ejecta-DOM model has $\kappa=0.03 ~{\rm cm^2/g}$, $\Delta t_{\rm exp}=5 \times 10^3 \s$, $M_{\rm DOM}=0.01 \msun$, $f_{\rm DOM}=0.15$ and the rest of the parameters as in section \ref{sec:method}.
The ejecta-companion model has a binary separation of $a = 2 \times 10^{12} \cm$ as in \cite{Dimitriadisetal2019a}.
We show this fit mainly to illustrate that the two interaction models have similar fitting power.
The shape of the interaction light curves is concave and unlike the observed linear rise as noted by \citet{Shappeeetal2018}.
The flux increases as the photosphere radius expands homologously, and the flux slope declines as the photosphere also recedes into the ejecta and the shock luminosity decreases with time.
Thus in any interaction model viewed in a wide optical band-pass the light curve shape is likely concave.

\begin{figure}[ht]
\includegraphics[scale=0.45]{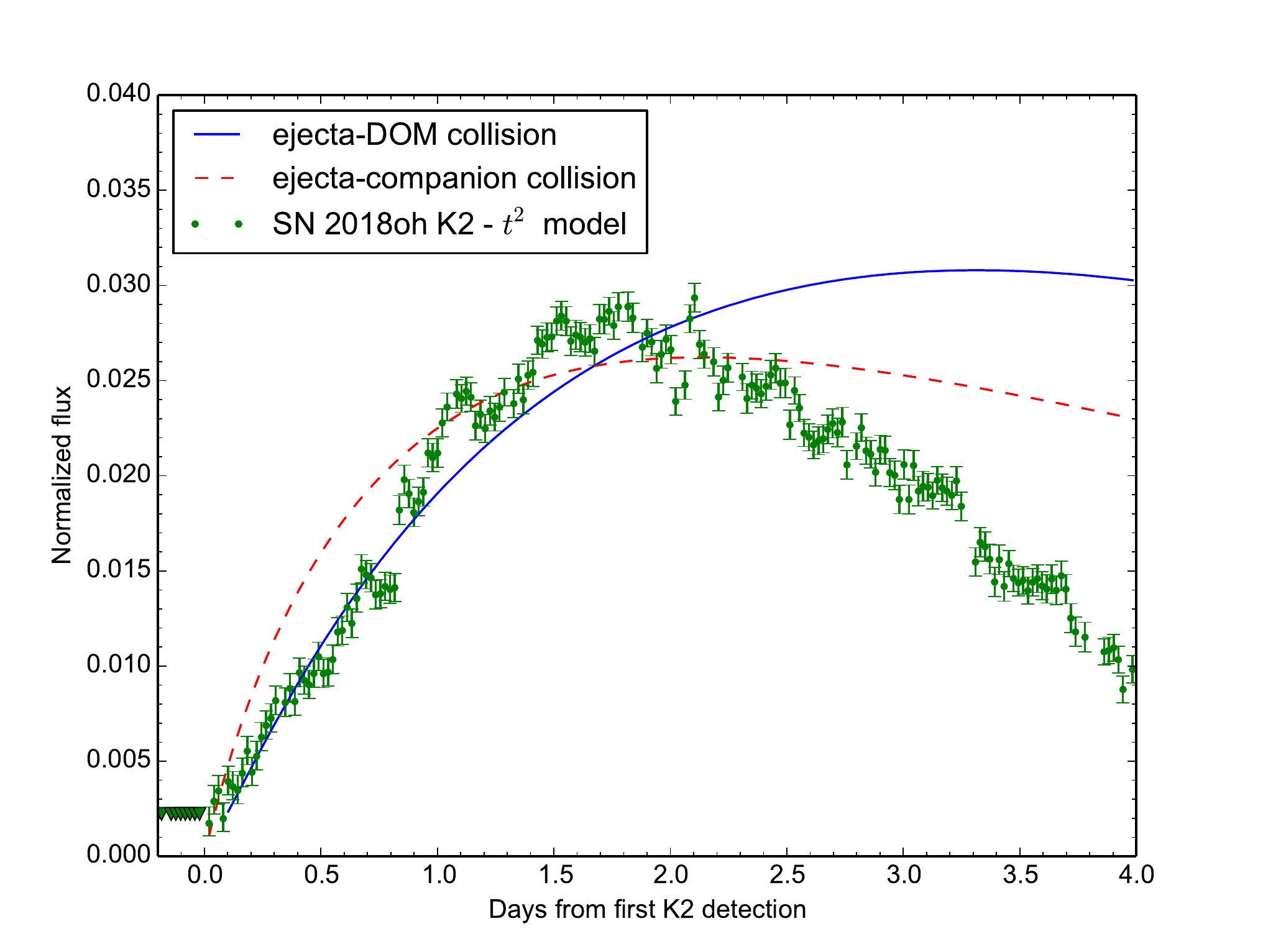}
\caption{Fitting an ejecta-DOM model to excess flux only.
Here excess flux is total flux minus a $L\propto t^2$ model for the contribution of the main nickel heating light curve.
Plot markings as in Fig. \ref{fig:dom-vs-sd-flux}.}
\label{fig:dom-flux-excess}
\end{figure}

Fig. \ref{fig:dom-flux-excess} shows the excess luminosity after subtracting a $L \propto t^2$ power-law fit from it.
We use the same range as \citet{Dimitriadisetal2019a} to fit a power-law model and subtract it from the K2 flux.
The ejecta-DOM model fit to the rising part of the excess light curve has $\kappa=0.2 ~{\rm cm^2/g}$, $\Delta t_{\rm exp}=5 \times 10^3 \s$, $M_{\rm DOM}=0.1 \msun$, $f_{\rm DOM}=0.1$ and the rest of the parameters as in section \ref{sec:method}.
\revision{The companion collision model has a separation of $a = 7 \times 10^{11} \cm$.}
Our model can only explain the rising part of the light curve as we do not model the gradual weakening of the shock in the ejecta behind the DOM.
The ejecta-companion model is also shown as in Fig. \ref{fig:dom-vs-sd-flux} for reference, though it was not meant to fit just the excess flux.

As we noted in section \ref{sec:method}, assuming the SN exploded before first light based on extrapolating a power-law is inconsistent with interaction models.
This means the above attempt to explain only the excess after subtracting the power-law model is also inconsistent.
We nevertheless show that it is possible to fit an ejecta-DOM collision model to it for completeness.

When claiming a light curve shows excess light we must specify what baseline is it relatively excessive to.
The above inconsistency is one reason we disfavour empirical power-law models for the underlying light curve, as we also discussed in our previous paper \citep{LevanonSoker2017}.
Instead of using an empirical model we tried to fit the analytic nickel heating models of \citet{PiroNakar2014} and find they cannot explain the first two days after first K2 detection.
\revision{This agrees with what \citet{Shappeeetal2018} and \citet{Dimitriadisetal2019a} find for SN~2018oh and what \citet{Maeda2018} find in general using more complete numerical models: interaction models dominate the light curve during the first $1-2 \days$ after explosion compared to nickel heating models without surface nickel.}
We do not assume an explosion model with surface nickel, and therefore consider the entire flux in the first $1-2 \days$ after first K2 detection as excessive.
\revision{For this reason we do not add a nickel heating component to the interaction models.}

\begin{figure}[ht]
\includegraphics[scale=0.45]{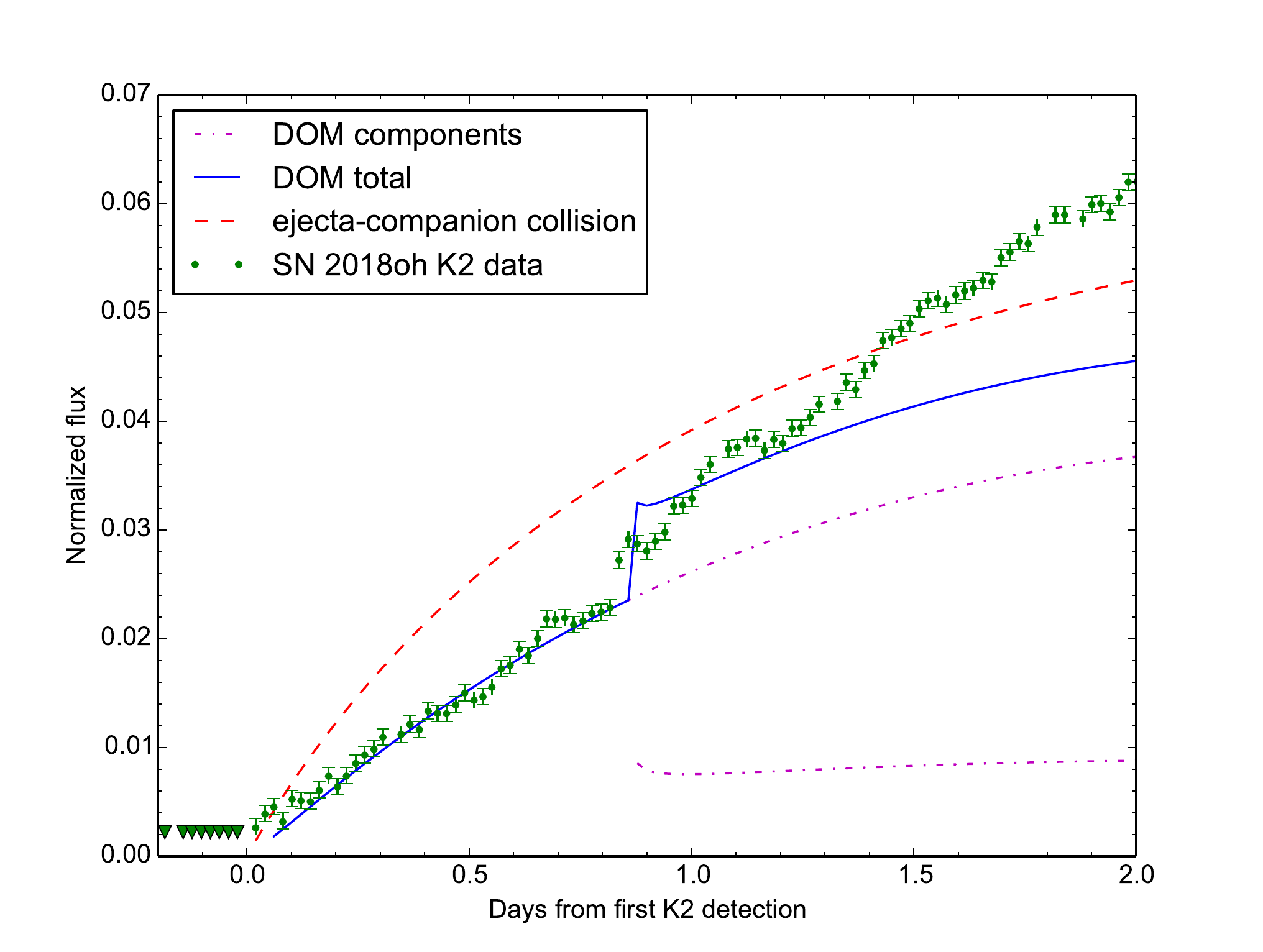}
\caption{Fitting the light curve with a two-component DOM model.
Plot markings as in Fig. \ref{fig:dom-vs-sd-flux}, with fluxes of each separate component in dot-dashed magenta.
The first DOM component (upper magenta line) has a velocity of $5000\kms$ and matches the earliest part of the light curve.
The second DOM component (lower magenta line) has a velocity of $10000\kms$ and the ejecta shocks it later than the first component, adding less flux at a later time compared to the first component.}
\label{fig:dom-flux-two-components}
\end{figure}

As we mentioned above, the DOM is likely to have a complicated bipolar structure rather than a one-component structure.
We here limit ourselves to examine only a two-component DOM structure \revision{as we draw schematically in the lower half of Fig. \ref{fig:schematic}.}
Fig. \ref{fig:dom-flux-two-components} shows a combination of two separate DOM components to explain the light curve up to later times.
The first DOM component has $\kappa=0.03 ~{\rm cm^2/g}$, $\Delta t_{\rm exp}=5 \times 10^3 \s$, $M_{\rm DOM}=0.01 \msun$, $f_{\rm DOM}=0.1$ and $v_{\rm DOM}=5000\kms$ (velocity as in the previous one-component DOM models).
The second DOM component has $\kappa=0.2 ~{\rm cm^2/g}$, $\Delta t_{\rm exp}=5 \times 10^3 \s$, $M_{\rm DOM}=0.05 \msun$, $f_{\rm DOM}=0.05$ and $v_{\rm DOM}=10000\kms$.
The rest of the DOM parameters are as in section \ref{sec:method}.
\revision{The same ejecta-companion model as in Fig. \ref{fig:dom-vs-sd-flux} is shown for reference.}
The two-component DOM model fits the observed light curve better than the one-component model.

The possibility of multiple DOM components adds another degree of freedom to ejecta-DOM collision models compared to an ejecta-companion model.
We emphasize that the DOM parameter configuration is degenerate.
Other sets of DOM parameters may yield similar results and the components are given here as an example.
However, the ability to explain long-lasting excess luminosity with multiple components is limited since at late times the photosphere is not expanding relatively as much as at early times.
This means that the flux slope at earliest times cannot be reproduced by having another DOM component shock the ejecta later at a greater distance.
Thus the additional degrees of freedom do not make the DOM model arbitrarily tunable to the rising light curve.

\section{Summary}
\label{sec:summary}

We addressed the early excess emission from the Type Ia SN~2018oh.
In our view there are three types of processes that might account for the excess emission.
These are the collision of the ejecta with a companion in the frame of the SD scenario, the collision of the ejecta with a DOM in the frame of the DD scenario, and the presence of radioactive nickel in the outskirts of the ejecta that can take place in all SN~Ia scenarios.
\revision{We summarise these processes in Table \ref{Table1}.}
\begin{table}[t]
    \begin{center}
    \caption{Early excess emission processes for SN~2018oh}
    \label{Table1}
    \begin{tabular}{c c c c} 
        \hline
                   & ejecta-companion   & ejecta-DOM         & $^{56}$Ni in       \\
                   & collision          & collision          & outer ejecta       \\
        [0.5ex]
        \hline
        \hline
        Scenario   & SD scenario        & DD scenario        & all                \\
        \hline
        Blue color & \greentext{\cmark} & \greentext{\cmark} & \redtext{\xmark}   \\
        \hline
        No H/He    & \redtext{\xmark}   & \greentext{\cmark} & \greentext{\cmark} \\
        \hline
    \end{tabular}
    \end{center}
\revision{Comparison of the three processes that might give an early emission excess in SN~2018oh.
The ejecta-companion collision predicts the presence of hydrogen and helium spectral lines that are not observed.
The presence of nickel in the outer ejecta predicts redder colors than those observed.}
\end{table}

\cite{Dimitriadisetal2019b} summarise their study of SN~2018a by stating that there are no known models that can simultaneously explain the blue early-time flux excess and the lack of late-time narrow emission lines.
However, they did not consider an ejecta-DOM interaction \citep{Levanonetal2015, LevanonSoker2017}.

In this study we showed that the ejecta-DOM interaction can fit the light curve as well as the ejecta-companion interaction or possibly better (Figs. \ref{fig:dom-vs-sd-flux}-\ref{fig:dom-flux-two-components}).
Since it does not suffer the drawbacks of the other processes, as we detail below, we consider the ejecta-DOM interaction to be the most likely explanation to the early excess emission in SN~2018oh.

The ejecta-companion interaction predicts the presence of hydrogen and/or helium in the late time spectrum, something that is not observed in SN~2018oh \citep{Tuckeretal2018, Dimitriadisetal2019b}.
The ejecta-DOM interaction does not predict any hydrogen or helium lines.

While the companion strongly shocks the ejecta in a well defined location and hence predicts too steep a rise \citep{Shappeeetal2018}, the DOM can spread over a greater volume and can shock the ejecta over a longer time in a gentler manner.
In Fig. \ref{fig:dom-flux-two-components} we present a toy model where we build a DOM from two components.
As evident from the figure, this toy model can better explain the rise in the first day compared to an ejecta-companion interaction.
A better fitting procedure for a multiple-component DOM structure may provide an even better fit of the light curve.
We presented a semi-analytic model for the ejecta-DOM collision.
A numerical simulation of DOM formation and collision will provide additional insights on this model.
This is the subject of a future hydrodynamical study.

\cite{Lietal2018} report the presence of prominent carbon absorption features in SN~2018oh that persist for an unusually long time.
We can speculate that some or all of this unburned carbon resides in the DOM before explosion, and mixes with the ejecta in the first hours after explosion as the ejecta hits and shocks the DOM.

We cannot conclude from our study that the DD scenario is the main SN Ia scenario, as in most cases observations do not include such an early epoch and interaction models are only seen from favourable viewing angles.
It is possible that in many cases the WD-WD interaction does not blow the DOM.
This leaves open the question of when the WD-WD interaction does blow a DOM and when it does not.
Our study does show that the DD scenario can best account for the early excess emission of SN~2018oh.
The DD scenario might be the main sub-Chandrasekhar SN scenario (e.g., \citealt{Maozetal2014}), including many WD-WD mergers from the hybrid channel, i.e., at least one of the WDs is a carbon-oxygen WD with large helium content (e.g., \citealt{Zenatietal2019}).
We look forward to the continuation of high cadence surveys with detection near first light and additional observations of unique early light curve behaviour -- perhaps not so unique after all.

\section*{Acknowledgements}

\revision{We thank Avishai Gilkis, Georgios Dimitriadis and an anonymous referee for helpful comments.}
This research was supported by the E. and J. Bishop Research Fund at the Technion and by a grant from the Israel Science Foundation.

\label{lastpage}
\end{document}